\documentclass[reprint,amsmath,amssymb,aps,prl,showpacs,floatfix,superscriptaddress]{revtex4-1}
\usepackage{color,graphicx,dcolumn}
\usepackage[colorlinks=true,urlcolor=blue,linkcolor=blue,
            citecolor=blue]{hyperref}

\begin{document}

\title{Quantum Monte Carlo study of the energetics of the rutile,
  anatase, brookite, and columbite TiO$_2$ polymorphs}

\author{John Trail}
  \email{jrt32@cam.ac.uk}
  \affiliation{Theory of Condensed Matter Group, Cavendish Laboratory,
               J.\ J.\ Thomson Avenue, Cambridge CB3 0HE, UK}

\author{Bartomeu Monserrat}
  \affiliation{Theory of Condensed Matter Group, Cavendish Laboratory,
               J.\ J.\ Thomson Avenue, Cambridge CB3 0HE, UK}
  \affiliation{Department of Physics and Astronomy,
               Rutgers University, Piscataway,
               New Jersey 08854-8019, USA}

\author{Pablo L\'opez R\'ios}
  \affiliation{Theory of Condensed Matter Group, Cavendish Laboratory,
               J.\ J.\ Thomson Avenue, Cambridge CB3 0HE, UK}

\author{Ryo Maezono}
  \affiliation{School of Information Science, JAIST,
               Asahidai 1-1, Nomi, Ishikawa 923-1292, Japan}

\author{Richard J.\ Needs}
  \affiliation{Theory of Condensed Matter Group, Cavendish Laboratory,
               J.\ J.\ Thomson Avenue, Cambridge CB3 0HE, UK}

\date{\today}

\begin{abstract}
  The relative energies of the low-pressure rutile, anatase, and
  brookite polymorphs and the high-pressure columbite polymorph of
  TiO$_2$ have been calculated as a function of temperature using the
  diffusion quantum Monte Carlo (DMC) method and density functional
  theory (DFT).
  The vibrational energies are found to be important on the scale of
  interest and significant quartic anharmonicity is found in the
  rutile phase.
  Static-lattice DFT calculations predict that anatase is
  lower in energy than rutile, in disagreement with experiment.
  The accurate description of electronic correlations afforded by DMC
  calculations and the inclusion of anharmonic vibrational effects
  contribute to stabilizing rutile with respect to anatase.
  Our calculations predict a phase transition from anatase to rutile
  TiO$_2$ at $630\pm210$~K.
\end{abstract}


\maketitle


Titanium dioxide (TiO$_2$) finds many applications in photocatalysis
\cite{Bai_TiO2_Chem_Rev_2014, Hashimoto_history_TiO2_2005} and as a
catalyst support \cite{Bourikas_catalyst_support_2014}.
It exhibits high chemical and optical stability, long-term
durability, corrosion resistance, low cost and non-toxicity.
Two main polymorphs of TiO$_2$ occur in nature, rutile and anatase
\cite{Hanaor_2011}, and a third polymorph, brookite, also exists at
ambient conditions \cite{Brookite_Meagher_1979}.
Several high-pressure forms of TiO$_2$ have also been synthesized, the
most stable of which is columbite (TiO$_2$-II)
\cite{Jamieson_TiO2_II_1969}.

Experiment and extrapolations from experimental data suggest
that rutile is the most stable phase from $0$ to above $1300$~K
\cite{Smith_2009}.
Anatase and brookite are observed to undergo irreversible phase
transitions to the rutile form at high temperatures \cite{Smith_2009,
Aoki_2015}, indicating the presence of high energetic barriers between
different polymorphs.
Although this could suggest that anatase and brookite are metastable
at all temperatures, the high energetic barriers make it difficult to
determine phase transition temperatures between polymorphs.
This suggests that accurate computations of the relative stabilities of
the polymorphs could provide valuable insights into the phase stability
of the TiO$_2$ system.

Relative energies of TiO$_2$ polymorphs have been calculated in
numerous first-principles density-functional-theory (DFT) electronic
structure studies \cite{Lazzeri_2001, Dompablo_2011, Vu_2012,
De_Angelis_2014, Muscat_2002, Beltran_2006, Landmann_2012,
Withers_2003, Zhu_2014}, which suggest that the three low-pressure
polymorphs, rutile, anatase, and brookite, are close in energy under
ambient conditions, although the results obtained depend significantly
on the density functional used, and are therefore inconclusive.

The main goal of the present work is to estimate the relative energies
of rutile, anatase, brookite and columbite TiO$_2$ with as high an
accuracy as possible.
We calculate the relative electronic energies of TiO$_2$ polymorphs
using the diffusion quantum Monte Carlo (DMC) method
\cite{Ceperley-Alder_1980, Foulkes_RMP_2001}, which is the most
accurate method known for calculating the energy of a large system of
quantum particles.
Furthermore, we find that the energetic contribution arising from
thermal nuclear motion, which we have calculated using DFT, is crucial
in determining the phase stability of TiO$_2$ polymorphs, and that
anharmonic terms are important in dynamically stabilizing the rutile
polymorph.
A very recent study has reported similar results, although anharmonic
vibrations were not considered \cite{Kent_TiO2}.


The quantum Monte Carlo calculations were performed using the
\textsc{casino} code \cite{CASINO_review}.
The cost of a DMC calculation scales approximately as the cube of the
number of particles $N$ for fermionic systems, which allows
applications to large systems.
The central approximation in a DMC calculation is the ``fixed-node
constraint'' \cite{Anderson_fixed_node_1976}.
The nodal surface is the ($3N-1$)-dimensional surface on which the
wave function is zero and across which it changes sign.
A trial wave function is optimized using the variational quantum
Monte Carlo (VMC) method, and the nodal surface of the DMC wave
function is constrained to equal that of the trial wave function.
An importance-sampling transformation is employed in DMC which ensures
that the important parts of the wave function are sampled most often,
and reduces the fluctuations in the energy.
In DMC the imaginary-time Schr\"odinger equation is used to evolve an
ensemble of electronic configurations towards the ground state
distribution.
VMC and DMC are variational techniques in the sense that they give an
energy that is higher than or equal to the exact energy, subject to a
small statistical error that can be systematically reduced by running
the simulations for longer.
The variational property of VMC and DMC aids cancellation of errors in
energy differences.

We used a trial wave function of Slater-Jastrow form
\begin{equation}
  \label{eq:slater-jastrow}
  \Psi_{\rm SJ}({\bf R}) \! =
    \exp[J({\bf R})]
    \det\left[ \psi_n({\bf r}_i^{\uparrow})\right]
    \det\left[ \psi_n({\bf r}_j^{\downarrow})\right] \;,
\end{equation}
where ${\bf R}$ denotes the positions of all of the electrons, ${\bf
  r}_i^{\uparrow}$ is the position of the $i$th spin-up electron,
${\bf r}_j^{\downarrow}$ is the position of the $j$th spin-down
electron, $\exp[J({\bf R})]$ is a Jastrow correlation factor
\cite{ndd_jastrow, General_Jastrow_2012}, and $\det{\left[ \psi_n({\bf
      r}_i^{\uparrow}) \right]}$ and $\det{\left[ \psi_n({\bf
      r}_i^{\downarrow}) \right]}$ are determinants of up- and
down-spin single-particle orbitals.
The DMC calculations were performed using the T-moves scheme which
ensures that the energy remains greater than the ground-state energy
when pseudopotentials are used \cite{Casula_T-moves_2006,
Casula_T-moves_2010, Drummond_2016}.

Single-particle orbitals for the TiO$_2$ structures were calculated
using the \textsc{castep} plane-wave DFT code
\cite{Clark_CASTEP_2005}, the PBEsol functional \cite{PBEsol_2008} and
a large basis-set energy cutoff of 160 Ry ($\simeq$ 2177 eV), which
provided DFT energy differences between structures converged to within
0.0036 eV/[TiO$_2$] of the large basis set limit.
The orbitals were transformed into a ``blip'' polynomial basis for
efficient evaluation in the calculations \cite{Alfe_blips_2004}.
We used Jastrow factors consisting of an electron-nucleus term
represented by a polynomial of order 8, an electron-electron term
represented by a polynomial of order 8, and a cosine expansion with 4
inequivalent parameters per spin-pair type, giving a total of 43
optimizable parameters.
The wave-function parameters were optimized using VMC, with the
variance of the energy minimized first \cite{ndd_newopt}, followed by
minimization of the variational energy \cite{umrigar_emin,
Umrigar_energy_optimization_2007}.


We used correlated-electron pseudopotentials (CEPPs) for O
\cite{CEPPS_sp_O} and Ti \cite{CEPPS_3d_Ti} generated from \textit{ab
initio} multi-configurational Hartree-Fock (MCHF) atomic calculations
\cite{Fischer_1,Fischer_2}.
The CEPPs give significantly more accurate results than Hartree-Fock
pseudopotentials \cite{Trail_smooth_relativistic_2005,
HF_pseudo_comparison_2014}.
CEPPs contain two-body operators that describe core polarization
\cite{CPP_Lee-Needs_2003} and their high accuracy has been verified in
tests using coupled cluster calculations including single, double, and
perturbative triple excitations [CCSD(T)]
\cite{CCSD(T)_Raghavachari_1989} with the \textsc{molpro} code
\cite{Werner_MOLPRO}.
The O CEPP has a He core and the $d$ channel was chosen as local.
A projector was generated from the ground state orbitals for both the
$s$ and $p$ channels.
The Ti pseudopotential has a Ne core and the $f$ channel was chosen as
local.
The semi-core nature of the Ti CEPP was captured using five
projectors.
We use the highest angular momentum channel available as the local
potential, which helps to reduce the errors associated with using
pseudopotentials in DMC \cite{Drummond_2016}.


The thermodynamic limit was approached in our calculations by using
large simulation cells containing 768 electrons (32 formula units)
subject to periodic boundary conditions.
We used experimental cell parameters and atomic positions at 300~K
from Refs.\ \onlinecite{Rutile_Swope_1995} (rutile),
\onlinecite{Anatase_Howard_1991} (anatase),
\onlinecite{Brookite_Meagher_1979} (brookite), and
\onlinecite{Columbite_Chen_2002} (columbite).
Cell dimensions, and the number of formula units contained within each
of them, are given in Table \ref{tab:geom}.
We chose simulation cells that maximize the distance between periodic
images \cite{Drummond_hydrogen_2015}, which mitigates finite-size
errors \cite{Drummond_finite_size_effects_2008,
Kent_finite_size_effects_1999}.

\begin{table*}[htb!]
\begin{tabular}{ldddcdd}
  \hline \hline \\[-0.33cm]
    & \multicolumn{3}{c}{Unit-cell lattice constants (\AA)}
    & formula
    & \multicolumn{2}{c}{$F(300~{\rm K})$ (eV/[TiO$_2$])} \\
  Polymorph
    & \multicolumn{1}{c}{$a$}
    & \multicolumn{1}{c}{$b$}
    & \multicolumn{1}{c}{$c$}
    & units
    & \multicolumn{1}{c}{\text{DMC}}
    & \multicolumn{1}{c}{\text{DFT}} \\
  \hline \\[-0.33cm]
  Rutile \cite{Rutile_Swope_1995}
    & 4.5922 & 4.5922 & 2.9574 & 2 & -2476.228(8) & -2470.2631 \\
  Anatase \cite{Anatase_Howard_1991}
    & 3.7845 & 3.7845 & 9.5143 & 4 & -2476.246(6) & -2470.3091 \\
  Brookite \cite{Brookite_Meagher_1979}
    & 9.174  & 5.449  & 5.138  & 8 & -2476.24(1)  & -2470.2826 \\
  Columbite \cite{Columbite_Chen_2002}
    & 4.61   & 5.43   & 4.87   & 4 & -2476.15(1)  & -2470.2254 \\
  \hline \hline
\end{tabular}
\caption{
  \label{tab:geom}
  Experimental orthorhombic unit cells used for the four TiO$_2$
  polymorphs.
  The Helmholtz free energies at 300~K, $F(300~{\rm K})$, estimated as
  the sum of electronic energy and the anharmonic vibrational free
  energy at 300~K are also shown.
  The anharmonic vibrational free energy is calculated using the PBEsol
  functional and VSCF equations, whereas the electronic energy is
  evaluated with both DMC and DFT using the PBEsol functional.
}
\end{table*}

DMC calculations were performed for each structure at a single
wavevector which was chosen from the wavevectors that yield a
real-valued wave function so as to minimize the difference between
the DFT energy at the single wavevector and the converged DFT energy.
We used the finite-size correction scheme of Kwee \textit{et al.\@}
\cite{Kwee_KZK_2008} in which an alternative local density
approximation (LDA) functional (the ``KZK'' functional, constructed
from DMC data for homogeneous electron gases) provides reference DFT
energies that include finite-size effects, which can then be used to
construct a finite-size correction for the DMC energies.

Our main results were obtained using a DMC time step $\tau$ of
0.004 a.u.\@ and a target population of 10,240 configurations.
While explicit time-step extrapolation is common practice
\cite{Lee_DMC_strategy_2011}, test DMC calculations with time steps of
$\tau = 0.001$, $0.0025$ and 0.004 a.u.\@ using simulation
cells containing 192 electrons (8 formula units) showed that
time-step errors in energy differences were negligible at $\tau =
0.004$ a.u., validating the use of a single time step.
More details of the finite-size corrections and extrapolations are
provided in the Supplemental Material \cite{Supplemental}.


Figure \ref{fig:static_lattice_energies} shows the energies of the
four polymorphs obtained in DMC and for DFT calculations using the
same set of functionals used to assess the accuracy of the DMC
finite-size corrections \cite{lda_1981, pbe_1996, PBEsol_2008,
pw91_1992, rpbe_1998, wc_2006}.
Energies are given relative to that of anatase for each method.
Calculations with ``on-the-fly'' \textsc{castep} DFT pseudopotentials
\cite{Clark_CASTEP_2005} produced similar results to those shown in
Fig.\ \ref{fig:static_lattice_energies}.

\begin{figure}[htb!]
  \includegraphics[scale=0.33]{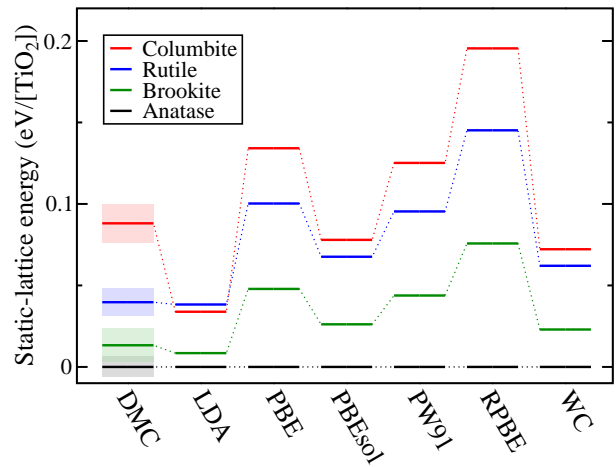}
  \caption{
    \label{fig:static_lattice_energies}
    Static-lattice energies relative to that of anatase from DMC and
    DFT calculations using the experimental lattice constants.
    The statistical errors in the DMC results are represented by
    translucent rectangles.
  }
\end{figure}

Given the relatively large variation of structural energies with the
choice of functional and the estimated errors for DMC results, these
data show that the DFT level of theory provides an inadequate
description of the relative energies of the four structures.
The generalized gradient approximation (GGA) functionals were found
to provide the same energetic ordering of the phases as DMC, but most,
with the notable exception of the PBEsol and WC functionals,
tend to overestimate the energy differences, while the LDA functional
significantly underestimates the static-lattice energy of columbite.
The relative DFT energies are expected to show significant
self-interaction errors which give rise to unphysical delocalization
of the $3d$ electrons.


The inclusion of nuclear vibrational motion provides significant
corrections to the static-lattice energies of TiO$_2$ polymorphs.
We have performed vibrational calculations for the four polymorphs at
the anharmonic vibrational level using a vibrational
self-consistent-field (VSCF) method \cite{Monserrat_anharmonic_2013}
and the PBEsol density functional.
The vibrational calculations were performed using the efficient
``nondiagonal supercells'' method
\cite{Lloyd-Williams_nondiagonal_supercells_2015} which allows us
to fully converge the results with respect to system size.

The inclusion of anharmonic contributions in our lattice dynamics
calculations is motivated by the presence of unstable vibrational
modes in rutile TiO$_2$ when described within the harmonic
approximation \cite{Mitev_soft_modes_2010, Lan_soft_modes_2015,
Wehinger_soft_modes_TiO2_2016}.
Using the experimental volume, we find unstable modes within various
regions of the vibrational Brillouin zone (BZ), mostly around the
$\Gamma$ point and the
$\left(\frac{1}{2},\frac{1}{2},\frac{1}{4}\right)$ point.
The presence of unstable modes is independent of the density
functional used to perform the calculations (see Supplemental
Material \cite{Supplemental}).
Instead of directly including anharmonic effects, first-principles DFT
calculations for TiO$_2$ have mostly been performed using structures
that are relaxed to volumes which are smaller than the experimental
volume, an approach that tends to remove the dynamical instabilities
but leads to a substantial bias in the energy differences between
structures.

\begin{figure}[htb!]
  \includegraphics[scale=0.33]{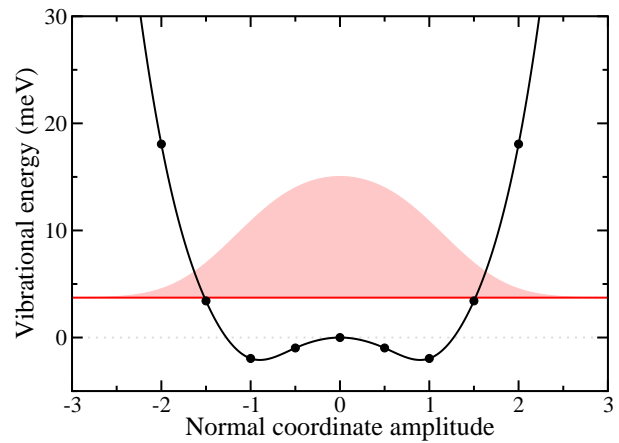}
  \caption{
    \label{fig:rutile_anharmonicity}
    A slice through the Born-Oppenheimer energy surface at the
    $\Gamma$ point of the Brillouin Zone.
    The red shaded area shows the vibrational density for a mode that
    is unstable at the harmonic level but is stabilized by
    anharmonicity.
    The horizontal red line shows the energy of the mode.
    The normal coordinate is measured in units of
    $1/\sqrt{2|\omega|}$, where $\omega$ is the imaginary harmonic
    frequency of the mode.
  }
\end{figure}

Figure \ref{fig:rutile_anharmonicity} shows a slice through the
Born-Oppenheimer (BO) energy surface of rutile TiO$_2$.
The atomic configurations at which we have evaluated the BO energies are
along a line spanned by one of the two soft modes at the $\Gamma$ point.
Our VSCF calculations show that the inclusion of anharmonicity leads to
dynamical stability throughout the BZ, arising primarily from quartic
terms in the BO potential.
Figure \ref{fig:rutile_anharmonicity} illustrates the quartic
anharmonicity of the BO energy surface and the resulting vibrational
density.
The explicit inclusion of anharmonicity plays a significant role in
the energetics of rutile TiO$_2$, with the anharmonic free energy
differing from harmonic estimates of the free energy by up to 0.035
eV/[TiO$_2$] (see Supplemental Material \cite{Supplemental}).

\begin{figure}[htb!]
  \includegraphics[scale=0.33]{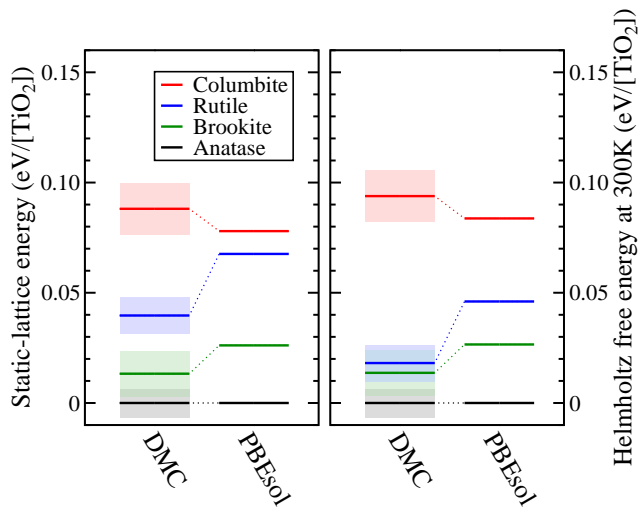}
  \caption{
    \label{fig:free_energy_300K}
    Left: static-lattice energies relative to that of anatase from DMC
    and DFT-PBEsol calculations, and right: Helmholtz free energies at
    300~K evaluated by adding DFT-based vibrational corrections to the
    DMC and DFT-PBEsol energies.
    The statistical error in the DMC results is represented by
    translucent rectangles.
    The DFT data were generated using a ${\bf k}$-point grid spacing
    of 0.060 \AA$^{-1}$ with an error estimated to be less than
    0.004 eV/[TiO$_2$].
  }
\end{figure}

We take the Helmholtz free energy to be the sum of the static-lattice
electronic energy and the anharmonic vibrational free energy at 300~K,
which is a valid approximation due to the large band gap of TiO$_2$
polymorphs relative to the thermal energy.
Figure \ref{fig:free_energy_300K} shows the total static-lattice
energies and the Helmholtz free energies at $300$~K using electronic
energies evaluated with DMC and DFT using the PBEsol functional.
The DMC Helmholtz free energies of anatase, brookite, and rutile are
indistinguishable within our target accuracy of 0.01 eV/[TiO$_2$].
Our final DFT and DMC Helmholtz free energies at $300$~K including the
effects of anharmonic vibrations are summarized in Table
\ref{tab:geom}.

\begin{figure}[htb!]
  \includegraphics[scale=0.33]{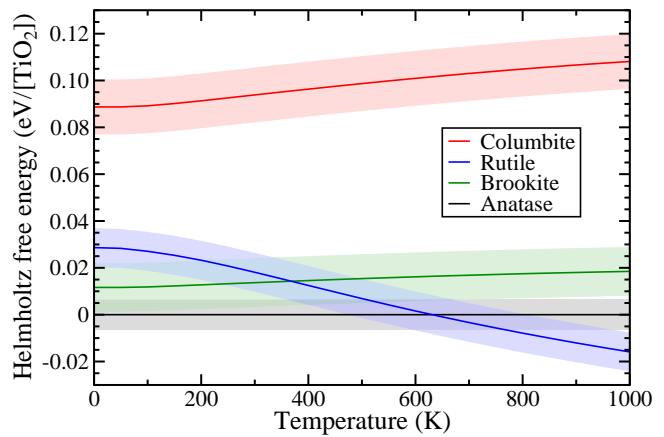}
  \caption{
    \label{fig:free_energy_vs_T}
    Helmholtz free energies relative to that of anatase as a function
    of temperature.
    These are estimated as the sum of the DMC electronic energy and
    the DFT anharmonic vibrational free energy at each temperature.
    Both the electronic energies and the anharmonic vibrational
    free energies are evaluated for 300~K experimental unit cells
    at all temperatures.
    Statistical errors are represented by translucent regions.
    Details of the DFT anharmonic vibrational energy calculations are
    as in Fig.\ \ref{fig:free_energy_300K}.
  }
\end{figure}

We also evaluate the Helmholtz free energy at a range of temperatures,
using the $300$~K experimental unit cells.
The resulting Helmholtz free energies, relative to anatase, are shown
in Fig.\ \ref{fig:free_energy_vs_T}, suggesting a phase transition
from anatase to rutile at $630\pm210$~K.
We estimate the error incurred by neglecting thermal expansion
in our calculations to be less than $0.003$ eV/[TiO$_2$] over the
temperature range $300$--$575$~K by comparing the PBEsol harmonic free
energies of anatase and rutile TiO$_2$ at the fixed-volume geometries
with those at the variable-volume experimental unit cell geometries
provided by Hummer {\it et al.\@} \cite{Hummer_2007}.

Our results using the experimental geometries at 300~K for all
temperatures predict the stability of anatase at low temperatures.
However, the experimentally observed rutile polymorph is energetically
very close to being stable, which strongly suggests that reproducing
the experimental results requires both an accurate description of
electronic correlation and anharmonic vibrations.

There are several sources of error in our calculations.
The DMC energies are affected by time-step and population-control
bias, although we have verified that these effects are negligible
compared with our target accuracy of 0.01 eV/[TiO$_2$].
We have made every effort to minimize the fixed-node errors, which we
expect to be small, although they are difficult to quantify.
We have employed finite-size corrections using methods that have been
demonstrated to be accurate and consistent with direct extrapolation
\cite{Drummond_hydrogen_2015}.
Similarly, while pseudopotentials are inherently approximate,
our tests have shown that CEPPs provide very accurate representations
of the Ti and O atoms.
The use of pseudopotentials in DMC incurs an additional bias, but
the T-moves scheme ensures that this bias is positive, which promotes
cancellation of errors in energy differences.

We suggest that the disagreement with experiment is likely to arise
mainly from the DFT evaluation of vibrational properties.
DFT calculations with both the LDA and PBE density functionals have
been reported to underestimate harmonic frequencies in diamond
\cite{Maezono_diamond_2007}, and the size of this underestimation
is of the order of magnitude of the difference between the Helmholtz
free energies of anatase and rutile TiO$_2$ at low temperatures
reported in our work.


In summary, we have obtained accurate estimates of the relative
energies of rutile, anatase, brookite, and columbite TiO$_2$ using DMC
methods, and of the relative Helmholtz free energies by combining
static-lattice DMC methods and anharmonic vibrational methods.
The main calculations were performed using DMC static-lattice
calculations, and finite-size corrections were obtained from the KZK
scheme.
Our results confirm that columbite TiO$_2$ is significantly higher
in energy than the three low-pressure phases.
The lowest (anatase) and highest (rutile) Helmholtz free energies of
the low-pressure phases at 300~K differ by only about 0.02
eV/[TiO$_2$], which is close to our target accuracy of 0.01
eV/[TiO$_2$].
We have shown that an accurate description of both electronic
correlation and vibrational energies are required to provide accurate
energetics for the TiO$_2$ polymorphs considered.
Our VSCF calculations show that the rutile structure of TiO$_2$ is
stabilized by quartic anharmonic vibrational motion.
Our results are consistent with the experimentally observed
irreversible phase transitions of anatase and brookite to the
rutile form at high temperatures, but do not explain the
stability of rutile at low temperatures
\cite{Smith_2009,Aoki_2015}.


\begin{acknowledgments}
  J.R.T., P.L.R., and R.J.N.\ acknowledge financial support from the
  Engineering and Physical Sciences Research Council (EPSRC) of the
  U.K.\ under grant number EP/J017639/1.
  B.M.\ acknowledges Robinson College, Cambridge, and the Cambridge
  Philosophical Society for a Henslow Research Fellowship.
  R.M.\ is grateful for financial support from MEXT-KAKENHI grants
  26287063, 25600156, 22104011, and a grant from the Asahi Glass
  Foundation.
  Computational resources were provided by the Archer facility of the
  U.K.'s national high-performance computing service (for which access
  was obtained via the UKCP consortium, grant number EP/K014560/1), by
  the Center for Information Science of the JAIST, and by the
  K-computer (supported by the Computational Materials Science
  Initiative, CMSI/Japan, under project numbers hp120086, hp140150,
  and hp150014).
  Preliminary calculations were performed using the Darwin
  Supercomputer of the University of Cambridge High Performance
  Computing Service (\url{http://www.hpc.cam.ac.uk}).
  Supporting research data may be freely accessed at [URL], in
  compliance with the applicable Open Data policies.
\end{acknowledgments}


\end{document}